\documentclass[11pt,preprint]{aastex}
\usepackage{emulateapj5}
\usepackage{onecolfloat}
\usepackage{natbib}
\usepackage{graphicx}

\newcommand{\mbf}[1]{\mbox{\boldmath{$#1$}}}

\shorttitle{Small scale stability of thermonuclear flames}
\shortauthors{R\"opke, Niemeyer, \& Hillebrandt}

\begin{document}

\title{On the small-scale stability of thermonuclear flames
in Type I\lowercase{a} supernovae}

\author{F. K. R\"opke\altaffilmark{1}, J. C. Niemeyer\altaffilmark{2}, and W. Hillebrandt\altaffilmark{3}}
\affil{Max-Planck-Institut f\"ur Astrophysik, Karl-Schwarzschild-Str. 1, 85741 Garching, Germany}

\altaffiltext{1}{fritz@mpa-garching.mpg.de}
\altaffiltext{2}{jcn@mpa-garching.mpg.de}
\altaffiltext{3}{wfh@mpa-garching.mpg.de}

\begin{abstract}
We present a numerical model which allows us to investigate
thermonuclear flames in Type Ia supernova explosions. The model is
based on a finite-volume explicit hydrodynamics solver employing PPM. Using the
level-set technique combined with in-cell reconstruction and flux-splitting
schemes we are able to describe the flame in the discontinuity
approximation. We apply our implementation to flame propagation in
Chandrasekhar-mass Type Ia supernova models.
In particular we concentrate on intermediate scales between the
flame width and the Gibson-scale, where the burning front is subject
to the Landau-Darrieus
instability. We are able to reproduce the theoretical prediction on
the growth rates of perturbations in the linear regime and observe the
stabilization of the flame in a cellular shape. The increase of the
mean burning velocity due to the enlarged flame surface is
measured. Results of our simulation are in agreement with semianalytical
studies.
\end{abstract}


\section{Introduction}

Owing to the vast range of relevant length scales involved in the
problem, it is impossible to resolve the full Type Ia supernova (SN Ia
in the following) explosion in multidimensional
numerical simulations. Therefore the task has to be tackled in
different approaches to gain insight to various aspects of the
explosion mechanism. 
Large Scale Calculations (LSCs) try to model the explosion on scales of
the stellar radius (for a recent example see \citet{reinecke2002b}).
The motivation of our work is to model
thermonuclear flames and to analyze the effective velocity of their
propagation which is a crucial input for LSCs.

For SN Ia various explosion mechanisms have been suggested (for a review see
\citet{hillebrandt2000a}).  We refer to the class of so-called  Chandrasekhar
mass models in which the C+O white dwarf accretes sufficient amounts
of material
to reach the Chandrasekhar mass $M_\mathrm{ch}$ where carbon burning
is initiated. At the prevailing
temperatures of typically around 
$10^{10} \, \mathrm{K}$ the thermonuclear burning is confined to a very
narrow region which is called a flame. Numerical simulations indicate
that a promising scenario is that the
explosion starts out as a deflagration which gets accelerated by
turbulence (``pure turbulent deflagration model''), or
makes a transition to a detonation later on (``delayed detonation model'').
Thus a main ingredient of all deflagration SN Ia
models is the acceleration of the burning velocity owing to turbulence
generated by instabilities of the flame. The wrinkling of the flame
front as a result of turbulent flow extends the surface area of the
front and thereby increases its mean propagation rate. 
On large scales the
dominating instability is the Rayleigh-Taylor instability, caused by
the stratification of dense fuel and lighter ashes inverse to the
gravitational field. Owing to this instability rising bubbles of
burning material form, leading to shear (Kelvin-Helmholtz)
instabilities on their surfaces. These effects are believed to
generate the necessary amount of turbulence to explain the burning
velocities required for SN Ia explosions. 

Nevertheless it remains crucial for the understanding of SN Ia
explosions to investigate the flame
evolution on intermediate scales, that is on scales much smaller
than the dimension of the exploding star, but still much larger than
the inner flame structure. The most important questions in this scale
range are: 
Is the turbulence in SN Ia entirely driven by large-scale
effects or can effects on intermediate scales contribute to the
increase of the turbulent burning velocity of the flame front? 
Is there
a mechanism to trigger a deflagration-to-detonation transition (DDT)
\citep{ivanova1974a,khokhlov1991a, woosley1994}? A possible effect
that could be the answer to both questions would be the existence of
so-called ``active turbulent combustion'' 
\citep{kerstein1996a,niemeyer1997b}.
Therefore interaction of flame instabilities
with a turbulent flow field is of particular interest. 

Here we report the first step to find an answer to these questions,
namely the study of the Landau-Darrieus (LD hereafter)
instability in a quiescent flow in two spatial dimensions.
The LD instability \citep{landau1944a, darrieus1938a} acts on
propagating density discontinuities such as thin chemical or
thermonuclear flame fronts. In the nonlinear regime, this instability
is known to give rise to the formation of a cellular structure that
stabilizes the front against further growth of perturbations
\citep{zeldovich1966a}. There is a
number of previous publications dealing with this
topic, e.g. \citet{gutman1990a}, \citet{blinnikov1996a},
\citet{helenbrook1999a}. Our approach is in the spirit of the work by
\citet{niemeyer1995a} who first demonstrated the existence of the LD
instability in thermonuclear burning in white dwarf matter at
densities relevant for SN Ia explosions in a numerical simulation. The main difference between their and our
approach lies in the flame description. \citet{niemeyer1995a} resolve the thermal structure of the flame
thereby restricting the spatial scale space that can be investigated
numerically. In contrast to their model we describe the flame in a
fully parametrized way as a
discontinuity making use of level set and in-cell
reconstruction/flux-splitting techniques. This method stands in direct
succession of \citet{reinecke1999a} and gives us more flexibility in
numerical experiments.

\section{Flame model}
\label{section_flame}

In general, the structure of laminar deflagration flames can be thought of as
composed of
two zones: the convective
diffusive zone where the fuel is preheated (in our case by electron
conduction) to reach non-negligible 
reaction rates and the thin reactive diffusive zone where
the reaction actually takes place. This picture sets both the width $d_l$
and the propagation velocity $s_l$ of the laminar flame with respect
to the unburnt matter. A study
of laminar flames in the context of thermonuclear burning in degenerate
matter by means of direct numerical simulation can be found in \citet{timmes1992a}. 

The burning releases energy and
heats the burnt material resulting in an increase of temperature
across the flame. In case of thermonuclear burning in white
dwarfs the fuel consists of material containing a degenerate electron
gas. Since the degeneracy of the relativistic electron gas
is partly removed in the burnt matter, the temperature increase across the
flame yields a decrease in density. Thus, if the scales under
consideration are much larger than the flame
width (less than a centimeter for laminar flames in C+O white dwarfs,
see \citet{timmes1992a}), it is well justified to simplify the
flame model to a moving discontinuity in the state variables.
This discontinuity approximation does
not resolve the inner structure of the flame. It is therefore
necessary to prescribe the laminar burning velocity of the front as an
additional parameter taken from direct numerical simulations.

Thermonuclear burning in SN Ia is characterized by turbulent
combustion and takes place in the so-called flamelet regime for the
part of the explosion where most of the energy is released. Here
turbulence is driven on large scales by the rising Rayleigh-Taylor
bubbles invoking a turbulent cascade. The scaling law of this cascade
is still controversial, it may simply follow Kolmogorov-scaling
or, as pointed out by \citet{niemeyer1997d}, Bolgiano-Obukhov
scaling. In any case, the velocity fluctuation $u'$ is a function of
the scale $l$ and \citet{niemeyer1997d} show that there exists a scale
between the size of
the largest turbulent eddies and the flame width on which the eddy
turnover time becomes comparable to the flame crossing time. Therefore
the flame dynamics below that scale is not
affected by large-scale fluctuations. In case of chemical flames the
cutoff scale is known as Gibson length $l_\mathrm{gibs}$.

This defines the range in length scales we address in our
intermediate flame structure investigation as
\begin{equation}
\label{microscale}
d_l \ll l \ll l_\mathrm{gibs}.
\end{equation}
Realistic values for these length scales at a density of $5 \times
10^7 \mathrm{g}\,\mathrm{cm}^{-3}$ are less than $1 \, \mathrm{cm}$
for the flame width \citep{timmes1992a} and about $10^4 \, \mathrm{cm}$
for the Gibson length according to \citet{niemeyer1998b}.

\section{The Landau-Darrieus instability}
\label{section_theory}

On the intermediate scales defined by (\ref{microscale}) the burning
front can be described as a laminar flame, which is unaffected by
turbulence on large scales and dominated by the LD instability.
This instability is caused by
the refraction of the streamlines of the flow on the density change over
the flame (approximated as a discontinuity). The fluid velocity
component tangential to the flame front is steady and mass
conservation leads to a discontinuity in the normal velocity
component. Consider a flame front that is perturbed from an originally
planar shape. Mass flux conservation leads to a broadening of the flow
tubes in the vicinity of a bulge of the perturbation. Thus the local fluid
velocity is lower than the fluid velocity at $\pm \infty$. Therefore
the burning velocity $s_l$ of the flame is higher than the
corresponding local fluid velocity and this leads to an increment of the
bulge. The opposite holds for recesses of the perturbed front. In this
way the perturbation keeps growing. By means of a linear stability
analysis Landau found for the growth rate of the perturbation
amplitude
\begin{equation}
\label{landau-disp}
  \omega_{\mathrm{LD}} = k s_l \frac{\mu}{1 + \mu}
  \left(\sqrt{1 + \mu - \frac{1}{\mu}} - 1\right),
\end{equation}
where $k$ denotes the perturbation wavenumber and $\mu =
\rho_u/\rho_b$ is the density contrast across the flame front.  

\begin{figure}[t!]
  \begin{center}
  \includegraphics[width=9cm]{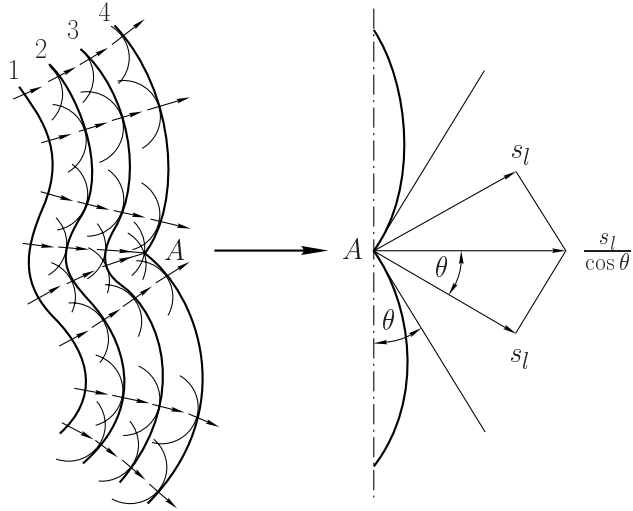}
  \end{center}
  \caption{Nonlinear stabilization of the flame front (adapted
  from \citet{zeldovichbook}).
  \label{zel}}
\end{figure}

The LD instability would in principle lead to unlimited growth of
small perturbations of the planar flame shape. It has, however,  been
ignored in all large scale SN Ia models
so far because there exists a nonlinear stabilization mechanism which
limits the perturbation growth.
Following \citet{zeldovich1966a} this effect can be explained in terms of
geometrical considerations: The
propagation of an initially 
sinusoidally perturbed flame is followed by means of Huygens'
principle (see Figure \ref{zel}).
After a while a cusp forms at the interfaces of neighboring
cells (marked with $A$ in Figure \ref{zel}). Here Huygens' principle breaks down and the flame propagation
enters the nonlinear regime. The propagation velocity at the cusp
exceeds $s_l$:
\begin{equation}
u_\mathrm{cusp} = \frac{s_l}{\cos \theta},
\end{equation}
where $\theta$ is the angle between the propagation velocity of the
cusp and the normal direction of adjacent cell segments.
This leads to a stabilization of the flame in a cellular
shape which has been observed in experiments. 

An analytical description of the flame propagation in the cellular
burning regime has been given by \citet{sivashinsky1977a} and by
\citet{frankel1990a}.  
The equations established there have been widely used in the chemical
combustion community and
were applied to the context of SN Ia by \citet{blinnikov1996a},
where a description of the cellular flame front in a fractal model is given. 
As the authors point out, these approaches to model flame fronts 
do not allow to
answer the question, weather the stabilization can break
up under certain conditions (e.g. certain densities; interaction with
turbulent velocity fields). In case of a break-up the LD instability
would lead to a
further increase of the effective burning speed. In this context a
transition from the deflagration mode of flame propagation to a
detonation (DDT) comes into consideration. A DDT has been successfully
implemented into empirical supernova models (\citet{hoeflich1996a},
\citet{iwamoto1999a}), but the
mechanism for it to occur 
is still unclear \citep{niemeyer1999a}.
Numerical simulations by \citet{niemeyer1995a} indicate a break-up
of the cellular stabilization and subsequent turbulization of the
burning front. One objective of the numerical model presented here is
a thorough investigation of this result. This is of great interest
because it could provide the mechanism for ``active turbulent
combustion'' (ATC) proposed by \citet{kerstein1996a} and discussed in
the context of SN Ia explosions by \citet{niemeyer1997b}.

In the following section we present a numerical method that will allow
us to study the stability of the cellular flame structure.

\section{Numerical Methods}
\label{section_nummeth}

Our implementation of the flame model goes back to the methods
described in \citep{reinecke1999a} where details of the numerical
implementation can be found. 
The fluid dynamics is described by the reactive Euler
equations. These are discretized on an equidistant cartesian grid and
solved by applying the Piecewise Parabolic Method
(PPM) suggested by \citet{colella1984a} in the PROMETHEUS
implementation \citep{fryxell1989a}.

The equation of state applied in our calculations comprises the
contributions that describe white dwarf matter. It includes an
arbitrarily degenerate and relativistic electron gas, local black body
radiation, an ideal gas of nuclei, and electron-positron pair
creation/annihilation (see e.g. \citet{cox1968a}).

The treatment of the flame in the discontinuity approximation is based
the so-called level-set technique that was introduced by
\citet{osher1988a}. The central idea of the level-set method is to
associate the flame with
a moving hypersurface $\Gamma(t)$ that is the zero level set of a
function $G(\mbf{r}, t)$:
\begin{equation}
  \Gamma(t) := \{ \mbf{r} \;|\; G(\mbf{r},t) = 0 \}.
\end{equation}
The function $G$ is prescribed to be a signed distance function
\begin{equation}
\label{distprop}
  |\mbf{\nabla} G| \equiv 1
\end{equation}
with
respect to the flame front and with \mbox{$G < 0$} in the unburned
material and \mbox{$G > 0$} in the ashes.
The temporal evolution of
$G$ is given by
\begin{equation}
\label{G_evolve_eqn}
  \frac{\partial G}{\partial t} = (\mbf{v}_u \mbf{n} + s_l)
  | \mbf{\nabla} G |
\end{equation}
for points located on the front. $\mbf{v}_u$, $\mbf
{n}$, and $s_l$
denote the fluid velocity in the unburnt region, the normal vector to
the flame front, and the burning speed with respect to the unburnt material,
respectively.

Our method crucially depends on the particular realization of the
coupling between the hydrodynamic flow and the flame which will be
reviewed in the following.

\subsection{Flame/flow coupling}
\label{sec_ffc}
In the context of the finite-volume method we apply
to discretize the
hydrodynamics, the cells cut by the flame front (``mixed cells'' in the
following) contain a mixture of
burnt and unburnt states. Therefore the quantity $\mbf{v}_u$ needed
in (\ref{G_evolve_eqn}) is not readily available.
One strategy to circumvent this problem is the so-called ``passive
implementation'' of the level-set method as described in
\citet{reinecke1999a}. There it is assumed that the velocity jump is
small compared with the laminar burning velocity and $\mbf{v}_u$ is
approximated by the average flow velocity. An operator splitting
approach for the time evolution of $G$ (\ref{G_evolve_eqn})  yields
the advection term owing
to the fluid velocity in conservative form which is identical to the
advection equation of a passive scalar. This part can be treated by
the PROMETHEUS implementation of the PPM method. Front
propagation, energy release, and species conversion as a result of burning are
accomplished in an additional step.

The strategy we use is called ``complete implementation'' by
\citet{reinecke1999a}. It was developed by
\citet{smiljanovski1997a}.
In these papers various tests of the numerical method, as for instance
the evolution of a spherical flame, are
presented. \citet{smiljanovski1997a} also compare with experimental
results for chemical flames. Our numerical implementation reproduces
the numerical tests successfully and because it does
not differ from the previous implementation in substantial points, we
will not repeat the tests here.

The special way of flame/flow coupling in the ``complete
implementation'' enables us to reconstruct the 
exact burnt and unburnt states in mixed
cells. This allows exact treatment of (\ref{G_evolve_eqn}). The main
advantage is that it becomes now possible to treat flows of burnt and
unburnt material over cell boundaries separately preventing the
flame front from smearing out over several cells as it does for the
``passive implementation''. The flame front is resolved as a sharp
discontinuity without any mesh refinement that would lead to very
small CFL timesteps. The ``complete implementation'' consisting of
in-cell reconstruction and flux-splitting schemes will be
reviewed in the following paragraphs. The description is restricted to
two-dimensional simulations.

\subsubsection{Geometrical information}

A prerequisite of the in-cell reconstruction and flux-splitting
schemes is the knowledge of geometrical quantities as the front
normal $\mbf{n}$, the unburnt volume fraction of mixed cell and the
unburnt fraction of the
interface area between the cells. The normal vector of the flame front
can be directly derived from the $G$ field,
\begin{equation}
  \mbf{n} = -\frac{\mbf{\nabla} G}{|\mbf{\nabla} G|},
\end{equation}
and is defined to point towards the unburnt material.

\begin{figure}[t!]
  \begin{center}
  \includegraphics[width=12cm]{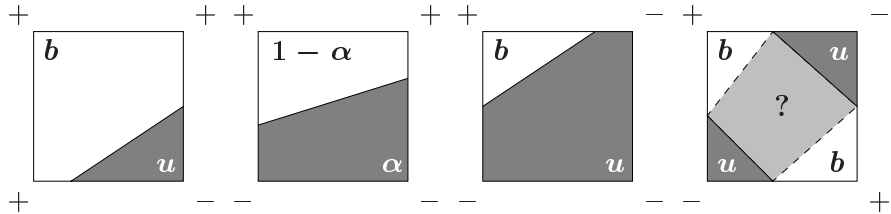}
  \end{center}
  \caption{Determination of $\alpha$ in mixed cells. The signs on the
  cell corners denote the sign of $G$ at these locations (adapted
  from \citet{reinecke1999a}).
  \label{alpha_fig}}
\end{figure}

The unburnt volume fraction $\alpha$ of a cell is obtained from the
intersections of the zero level set of the $G$-function with the cell
interfaces. Connecting these intersection points with a straight
line results in a piecewise linear approximation of the flame in
the cells and $\alpha$ is obtained by calculating the area behind the
connecting line. This procedure is depicted in
Figure \ref{alpha_fig}.
The rightmost sketch shows a situation where the
geometry is ambiguous. In these (fortunately rare) cases that
$\alpha$ of the two possibilities is taken which is closest to
the value of the preceding time step.

\subsubsection{In-cell reconstruction and flux-splitting}
\label{in-cell_rec_sec}
In order to reconstruct the unburnt and burnt states in mixed cells we
set up a system of equations.
The first three equations of that reconstruction system of equations
describe the cell averages of the conserved quantities as linear
combinations of the unburnt and the burnt part of the cell:
\begin{eqnarray}
  \overline{\rho} &=& \alpha \rho_u + (1 - \alpha) \rho_b
  \label{averagemass_eq}\\
  \overline{\rho\mbf{v}} &=& \alpha \rho_u \mbf{v}_u + (1 - \alpha) \rho_b
  \mbf{v}_b \label{averagemomentum_eq}\\
  \overline{\rho e} &=& \alpha \rho_u e_u + (1 - \alpha) \rho_b e_b
  \label{averagetotenergy_eq}
\end{eqnarray} 
Here, $\rho$ stands for the mass density, $\rho \mbf{v}$ for the
momentum density, and $\rho e$ for the
density of the total energy. The indices $u$ and $b$ denote the
unburnt and burnt quantities respectively.

Continuity of mass flux density, momentum flux density and energy
flux density over the flame front yield the Rankine-Hugoniot jump
conditions and the Rayleigh criterion. To incorporate the former into
the system of equations it is convenient to split the velocity vector into
a normal part $\mbf{v}_n$ and a tangential part $\mbf{v}_t$ with
respect to the front. Eq.~(\ref{averagemomentum_eq}) then reads
\begin{equation}
  \overline{\rho v}_n = \alpha \rho_u v_{n,u} + (1 -
  \alpha) \rho_b v_{n,b} \label{averagemomentumnormal_eq}
\end{equation}
and
\begin{equation}
  \overline{v}_t = v_{t,u} = v_{t,b}.
\end{equation}
The Rayleigh criterion and the Hugoniot jump condition for the
internal energy $e_i = e - \mbf{v}^2 / 2$ yield
\begin{equation}
  \label{rayleighcriterion_eq}
  (\rho_u s_l)^2 = \frac{p_u - p_b}{V_b - V_u}
\end{equation}
and
\begin{equation}
  \label{hugoniotcond_eq}
  e_{i,b} - e_{i,u} = \Delta w_0 - \frac{1}{2}(p_b + p_u)(V_b - V_u),
\end{equation}
respectively, with $V := 1/\rho$ and $\Delta w_0$ denoting the
difference between the formation enthalpies of ashes and fuel. The
pressures are given by the equation of state
\begin{equation}
\label{presseos_eq}
  p_u = p_\mathrm{EOS}(\rho_u, e_{i,u}, \mbf{X}_u); \quad 
  p_b = p_\mathrm{EOS}(\rho_b, e_{i,b}, \mbf{X}_b),
\end{equation}
where $\mbf{X}$ denotes the mass fractions of the species.
The jump condition for the normal velocity component must hold:
\begin{equation}
  \label{normalveljump_eq}
  v_{n,b} - v_{n,u} = s_l \left( 1- \frac{\rho_u}{\rho_b} \right). 
\end{equation}
Additionally, a burning rate law prescribing the laminar burning
velocity (e.g. in terms of the front geometry) must be provided. The
reconstruction system of equations consists of (\ref{averagemass_eq}),
(\ref{averagetotenergy_eq}), (\ref{averagemomentumnormal_eq}),
(\ref{rayleighcriterion_eq}), (\ref{hugoniotcond_eq}),
(\ref{presseos_eq}), and (\ref{normalveljump_eq}). Assuming the
composition of the fuel and the ashes as known, we end up with eight
equations for the unknown quantities $\rho_u$, $\rho_b$, $v_{n,u}$, $v_{n,b}$,
$e_{i,u}$, $e_{i,b}$, $p_u$, and $p_b$. Thus the nonlinear
reconstruction system of equations is closed and can be solved
iteratively.

A flame generates source terms in species and
energy which have to be treated in addition. Taking into account the
species source term $\omega_X$ the
mass fraction evolution reads
\begin{equation}
\label{species_evol}
\frac{\partial}{\partial t} (\rho X) + \mbf{\nabla} \cdot (\rho X
\mbf{v}) = \rho \omega_X.
\end{equation}
\citet{smiljanovski1997a} introduce a method for explicit
evaluation of the species source term. However, \citet{schmidt2001a}
points out that this method can raise severe complications which
cause a failure of the reconstruction. The reason is that the
species evolution is solved in a conservative way whereas the
$G$-evolution
is non-conservative leading to a discrepancy between the front shape
represented by the zero level-set of $G$ and the species
field. \citet{schmidt2001a} suggests an algorithm for an implicit
evaluation of the reaction term with help of the $G$-function leading
to a ``predictor'' of the energy in the hydrodynamics module which is
then corrected after the reconstruction. We follow this suggestion in
our implementation diverging from \citet{reinecke1999a}.

To solve the reconstruction system of equations we reduce it to four
equations in the unknown variables $\rho_u$, $\rho_b$, $e_u$, and
$e_b$ and employ the MINPACK
implementation \citep{more1980a} of Powell's hybrid scheme
\citep{powell1970a}.

After this reconstruction step we are able to compute the
hydrodynamical fluxes of conserved quantities over the cell interfaces
separately for the unburnt and burnt material. This procedure
guarantees that the newly computed mixed
state is consistent with the volume parts $\alpha$ and $(1 - \alpha)$
to be computed in the next time step. It also prevents the
burning front from smearing out over several computational
cells. Details on this technique can be found in \citet{reinecke1999a}.

\subsection{Thermonuclear reactions}
The thermonuclear burning considered in the presented simulation takes
place at fuel densities above $10^7$ g cm$^{-3}$ and terminates in
nuclear statistical equilibrium (NSE) which consists mainly of nickel.
Because of the high computational costs of the implementation of a
full nuclear reaction network and owing to the fact that our intention
is a the investigation
of flame dynamics rather than a correct nucleosynthetic description of
SNe Ia, we simplify the nuclear processes to an effective reaction:
$$
  14 \; ^{12}\mathrm{C} \longrightarrow 3 \; ^{56}\mathrm{Ni}.
$$ 
This yields a specific energy release of $q = 9.28667 \cdot 10^{17}$ erg
g$^{-1}$ \citep{audi1995a}. Consequently we model the initial
composition as being pure carbon.

\section{Results}

We present some basic studies
performed with the numerical implementation as described in
{\S} \ref{section_nummeth}. These investigations concern the
development of 
perturbations under the influence of the LD instability, the
transition to the nonlinear regime of flame propagation, and the
formation of a cellular flame structure as predicted by theory. We
will give a survey of results referring to an exemplary
fuel density of $5 \times 10^7 \mathrm{g}\,\mathrm{cm}^{-3}$ which was
chosen for our demonstration calculations and add some remarks about
critical issues of our implementation.

\subsection{Evolution of the flame front}

The ``experimental setup'' for the studies of the evolution of the
flame subject to the LD instability is the following:
The physical domain was set to be periodic in $y$-direction. On
the left boundary of the domain an outflow condition was enforced and
on the right boundary we imposed an inflow condition with the unburnt
material entering with the laminar burning velocity $s_l$. This would
yield a computational grid comoving with a planar flame. Since the LD
instability leads to a growth of the perturbation and therefore
increases the flame surface speeding up the flame, it is necessary to
take additional measures to keep the mean position of the flame in
the center of the domain.

A sinusoidal initial perturbation was imposed on the flame
shape. The state variables were set up with values for the burnt and
unburnt states taken from from one-dimensional simulations performed
with the ``passive implementation'' of the level-set method by
\citet{reinecke1999a}, imposing a given value for the density of the
fuel. For a fuel density of $5 \times 10^7
\mathrm{g}\,\mathrm{cm}^{-3}$ the initial values read:
\begin{itemize}
\item laminar burning velocity: $s_l = 8.74 \times 10^5 \mathrm{cm} \, 
\mathrm{s}^{-1}$ following \citet{timmes1992a}
\item density of fuel: $\rho_u = 4.99 \times 10^7
\mathrm{g}\,\mathrm{cm}^{-3}$
\item density of ashes: $\rho_b = 2.07 \times 10^7
\mathrm{g}\,\mathrm{cm}^{-3}$
\item internal energy of fuel: $e_u = 3.58 \times 10^{17}
\mathrm{erg}\,\mathrm{g}^{-1}$
\item internal energy of ashes: $e_u = 1.05 \times 10^{18}
\mathrm{erg}\,\mathrm{g}^{-1}$.
\end{itemize}
These values also served as initial guesses for solving
the reconstruction equations.

\begin{figure}[t!]
  \begin{center}
  \includegraphics[width=9cm]{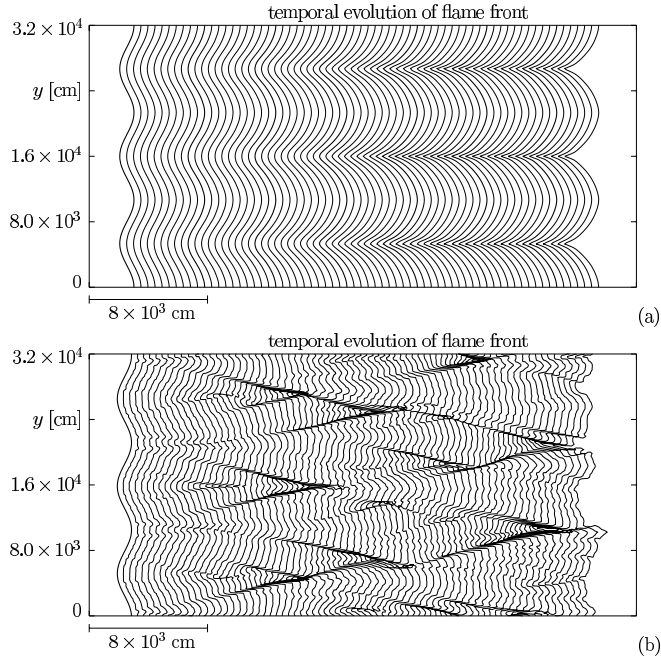}
  \end{center}
  \caption{Evolution of the flame front at  $\rho_u=5\times 10^7
    \mathrm{g}\,\mathrm{cm}^{-3}$ for a resolution of $200 \times 200$
   cells
   (a) in the ``complete implementation'';
   (b) in the ``passive implementation''.
 Each contour represents a timestep of 1.5 ms. 
  \label{evo_comp}}
\end{figure}

To demonstrate the superiority of the ``complete implementation'' of
the flame/flow coupling (see {\S} \ref{sec_ffc}) over the ``passive
implementation'' we compared the flame evolution for
a fuel density of  $5 \times 10^7 \mathrm{g}\,\mathrm{cm}^{-3}$ for both
methods. The result is shown in Figure \ref{evo_comp}. 

It illustrates very
drastically that the ``passive implementation'' fails to reproduce
theoretical expectations like an increase of the perturbation
amplitude due to the LD instability or the stabilization of the flame
in a cellular pattern. Both effects are present in the ``complete
implementation''. This difference can be attributed to the incorrect
treatment of the flame propagation velocity in the $G$-evolution
equation (\ref{G_evolve_eqn}) in the ``passive implementation''
preventing the flow field from adjusting to the perturbed flame shape,
which would cause the LD instability. It is evident that the ``complete
implementation'' is the appropriate tool for our study. All
simulations presented in the following are performed using this method.

\begin{figure}[t!]
  \begin{center}
  \includegraphics[width=18cm]{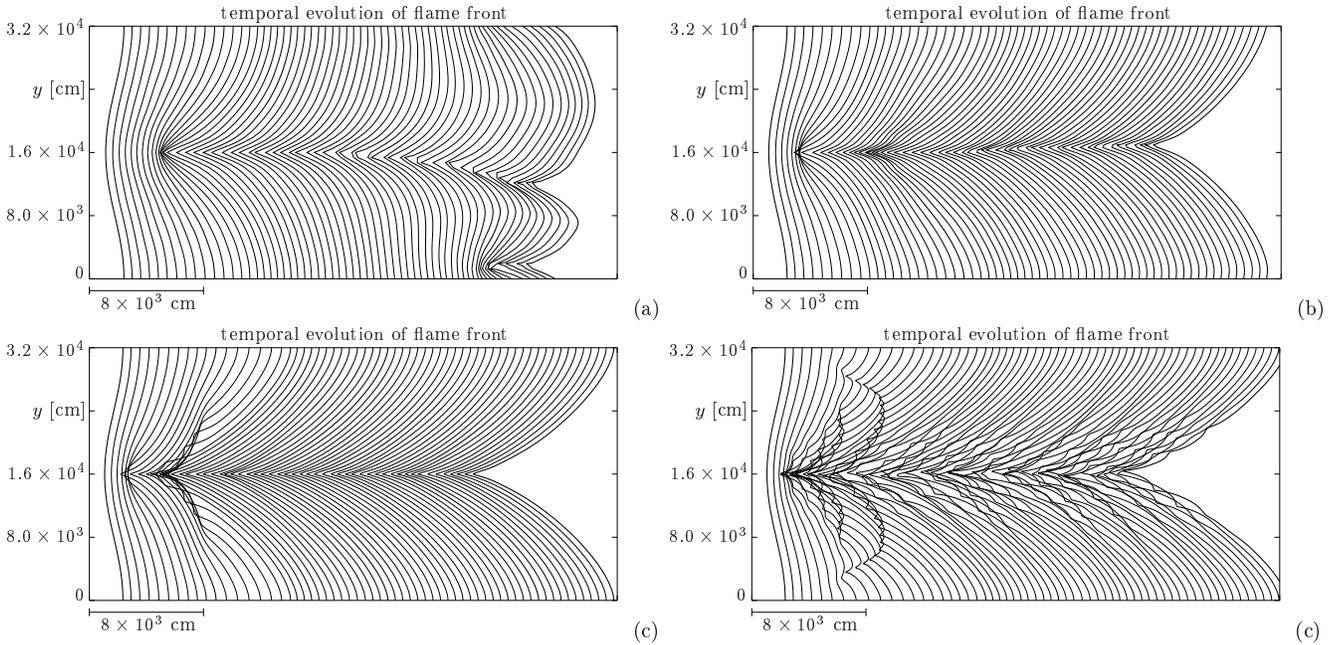}
  \end{center}
  \caption{Evolution of the flame front at  $\rho_u=5\times 10^7
    \mathrm{g}\,\mathrm{cm}^{-3}$. The
    contours mark evolution steps of 2.5 ms;
   (a) resolution 50$\times$50 cells;
   (b) resolution 100$\times$100 cells;
   (c) resolution 200$\times$200 cells;
   (d) resolution 400$\times$400 cells;
  \label{evo}}
\end{figure}

Figure \ref{evo} shows again the temporal evolution of a
flame front for a fuel density of $5 \times 10^7
\mathrm{g}\,\mathrm{cm}^{-3}$. The corresponding density contrast
across the flame is $\mu = 2.41$. This time the wavelength of initial
perturbation was chosen domain-filling. Five different resolutions of the
computational grid were selected: $50 \times 50$ cells, $100 \times 100$
cells, $200 \times 200$ cells, $300 \times 300$ cells (not shown in
Figure \ref{evo}), and $400 \times
400$ cells. The size of the domain amounted to
$\left[3.2\times10^4\mathrm{cm}\right]^2$. 
The planar
flame was
initially perturbed in a sinusoidal way with an amplitude of 640
cm.  But note that the length
scales are arbitrary since the problem is scale-invariant. In the
plots each contour is shifted
artificially into the $x$-direction for better visibility and
corresponds to a time evolution of 2.5 ms. The
$x$-coordinate is stretched by a factor of about 2.
 In all cases the initial perturbation grows and a cellular flame
shape emerges. For comparison we also
computed the propagation of a initially planar flame front in the same
setup and with a resolution of $400 \times 400$ cells over the same
time period. We found no sign of any deviation of the flame shape from
its initial planar configuration.

\subsection{The Landau-Darrieus instability}  
\label{sec_num_ld}

\begin{figure}[t!]
  \begin{center}
  \includegraphics[width=12cm]{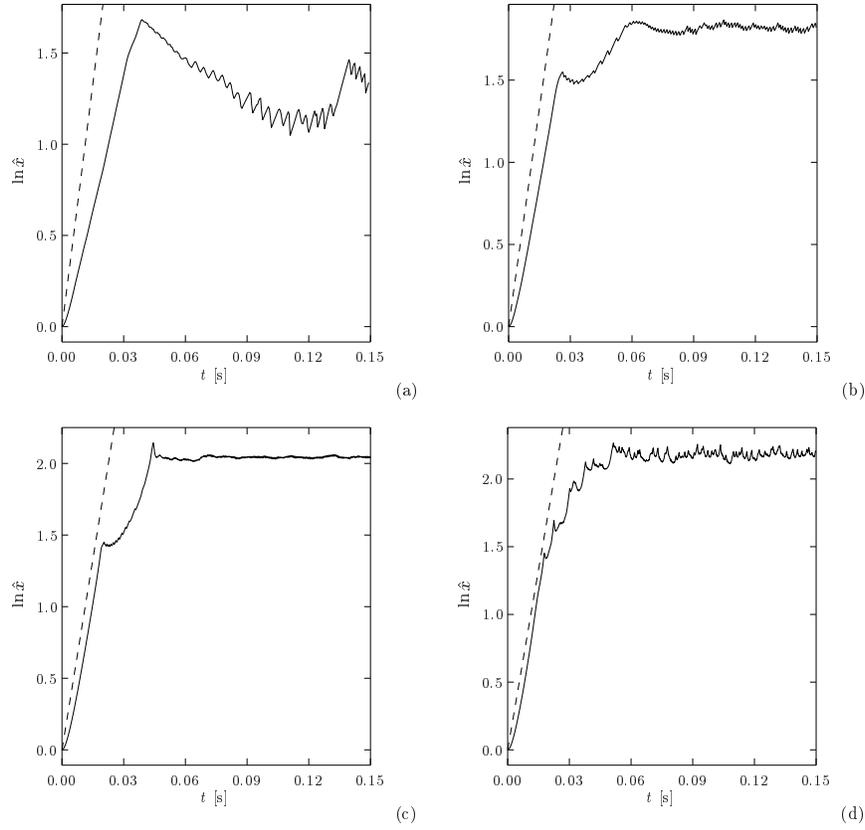}
  \end{center}
  \caption{Perturbation amplitudes $\ln \hat{x}$ as a function of time $t$;
   (a) resolution 50$\times$50 cells;
   (b) resolution 100$\times$100 cells;
   (c) resolution 200$\times$200 cells;
   (d) resolution 400$\times$400 cells;
  \label{amp}}
\end{figure}

Our first task was to measure the growth of the amplitude of the
perturbation and to compare it to the theoretical prediction
(eq.~(\ref{landau-disp})). This was done by a simple determination of the
distance between the rightmost and the leftmost points of the flame
front. Figure \ref{amp} shows the growth 
rate of the perturbation amplitude over time. The dashed line
corresponds to Landau's dispersion relation. All results share the
feature that the growth of the amplitude is shifted in the
beginning. This comes from the fact that we impose an initial
perturbation but not the corresponding flow field. Thus some
set-up period is necessary for the correct velocities to build up in
the vicinity of the flame front. In Figure \ref{amp}, the linear
regime of perturbation growth can be clearly distinguished from the
later nonlinear stage, where the perturbation amplitude saturates. We
will now discuss the linear part of the flame evolution.

It is evident from Figure \ref{amp} that the
difference between the growth of the perturbation amplitude in our
simulation and in theory decreases with resolution. According to Table
\ref{omega_tab} the deviation is about 46\% for the $50\times 50$
cells simulation and only about 10\% for the  $200\times 200$
cells simulation. In the higher resolved runs the initial perturbation
quickly gets superimposed with perturbations of higher wavenumber so
that the values of the growth rate can not safely be assigned to the
largest wavelength of perturbation anymore. This is the reason why it
exceeds the theoretical expectation slightly. One could conclude from
the results that the simulated growth rate matches the theoretical
expectations better with higher resolution and converge for a
resolution between 200 and 300 cells per dimension for our given fuel
density. Alternatively an explanation could be given in terms of a
finite effective flame thickness. Even with the special measures
described in {\S} \ref{section_nummeth} $d_l$ can not be expected to be
smaller than one cell width. A theoretical approximation of a finite
flame width is the assignment of a curvature-dependent flame speed:
\begin{equation}
\label{markstein}
s = s_l (1-l_{\mathrm{M}} \kappa),
\end{equation}
Here $\kappa$ denotes the curvature of the flame front and
$l_{\mathrm{M}}$ is the phenomenological Markstein length that is
related to the flame width.
This burning law changes the dispersion
relation in dependence of the Markstein length. In our case one would
expect the Markstein length to be higher for less resolved
simulations. The question whether the Markstein relation is
appropriate to describe the behavior of our flame front is subject to
a different study which will be presented
elsewhere.
\begin{deluxetable}{ll}
\tabletypesize{\scriptsize}
\tablecaption{Growth rate of perturbation amplitude\label{omega_tab}}
\tablewidth{0pt}
\tablehead{
\colhead{Case} & \colhead{$\omega$}
}
\startdata
Theory & 88.7\\
$50 \times 50$ cells & 47.8\\
$100 \times 100$ cells & 69.0\\
$200 \times 200$ cells & 79.7\\
$300 \times 300$ cells & 90.0\\
$400 \times 400$ cells & 91.4
\enddata
\end{deluxetable}

The sequence of plots in Figure \ref{amp} shows increasing similarity of
the overall evolution of the amplitudes with higher resolution. This
indicates that our numerical model 
of the LD instability in thermonuclear flames converges in the global
properties.

\subsection{Cellular stabilization}

Figure \ref{evo} clearly shows the transition in flame
propagation from the linear regime in the beginning, where perturbation
growth by virtue of the LD instability dominates the dynamics, to the
nonlinear burning regime. In accord with theoretical expectations
({\S} \ref{section_theory}) this happens by the formation of cusps in
recesses of the front. The flame adopts a cellular shape. 

The global picture of the 
cellular regime is similar in all resolutions but the
evolution of the shape differs in details. This difference is most prominent in
the lowest resolved flame where the cell splits and a two-cell
structure forms. As reported by \citet{joulin1994a} the phenomenon of
cell splitting is not supported by
analytical investigations of the nonlinear regime, but this could be
owing to the restricted class of solutions studied there. Its invocation
is usually attributed to numerical noise. This would explain why this
feature is observed in the low-resolution simulation where the
discretization errors are large.

All other runs share the
feature that the final outcome is a single-cell structure which
steadily propagates forward.
An effect contrary to the deviation in low resolved simulations is that
with increasing resolution perturbations of smaller wavelength become
observable that superimpose the initial long-wavelength
perturbation. This is evident for the highest resolved run. The
small cells superimposed to the large
cell propagate downward to the cusp where they disappear.
Again, the invocation of this phenomenon is discussed controversially
in literature and \citet{joulin1994a, joulin1989a} argues that it
should be attributed to numerical noise
(as roundoff and truncation
errors). It is even present in highly accurate numerical solutions of
the Sivashinsky-equation. However, a prerequisite for its observation
is a wide
enough computational domain (in our case a sufficiently high resolution in
$y$-direction), as reported by \citet{gutman1990a}.
The reason why a superimposed smaller-wavelength cellular pattern does
not destroy the global cusp-like structure although its amplitude
should grow faster due to the higher wavenumber (see eq.~(\ref{landau-disp})) was
explained by \citet{zeldovich1980a} on the basis of a WKB-like linear
analysis and by \citet{joulin1989a} in the nonlinear regime. The
origin of the effect lies in a flow component parallel to the flame
with increasing velocity towards the cusp. Therefore the
superimposed perturbation is advected towards the cusp while its
wavelength is stretched. In that way their grows is reduced until they
disappear in the cusp. This is exactly what we observe in our
highest-resolved simulation.

\subsection{Increase in flame surface and acceleration of the flame}

\begin{figure}[t!]
  \begin{center}
  \includegraphics[width=18cm]{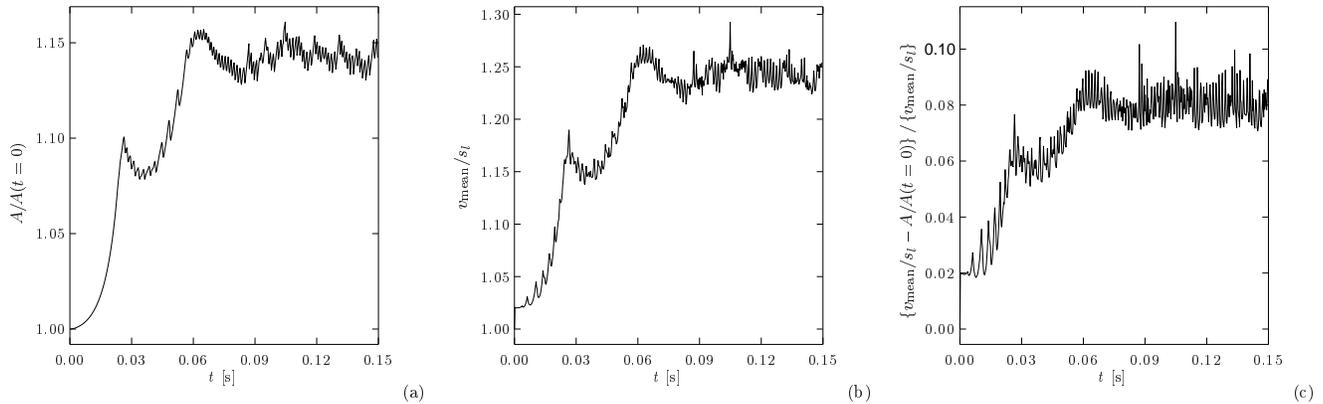}
  \end{center}
  \caption{Growth of flame surface area (a) and mean velocity (b) for
    a resolution of 100$\times$100 cells; (c) deviation of area and velocity
  \label{va}}
\end{figure}

The wrinkling of the flame front amplified by the LD instability
increases the flame surface. This causes the mean fuel consumption
rate to grow and the flame is expected to propagate with an
increased mean velocity $v_\mathrm{mean}$,
\begin{equation}
\label{area-vel}
\frac{v_\mathrm{mean}(t)}{v_\mathrm{mean}(t_0)} = \frac{A(t)}{A(t_0)},
\end{equation}
where $A(t)$ is the flame surface area (i.e. length of the
one-dimensional flame in our two-dimensional simulation) at time $t$.
In Figure \ref{va} we present a measurement of the
temporal evolution of the flame surface area and the mean
velocity. As expected, the evolution of the velocities follows the
growth of the initial perturbations in the beginning and saturates in
the cellular regime. 
The flame area in Figure \ref{va} is normalized to the
flame area of our initial configuration whereas the mean velocity is
normalized to the laminar flame speed. This results in  a shift between
$v_\mathrm{mean}(t) / s_l$ and ${A(t) / A(t_0)}$ because the initial
configuration is already perturbed from the planar shape and the
corresponding value of the mean velocity deviates from $s_l$. The flow
field needs some time to adapt to the initial flame
geometry. Figure \ref{va} (c) depicts the difference between
$v_\mathrm{mean}(t) / s_l$ and ${A(t) / A(t_0)}$. Apart from the
initial shift there is a deviation which can partly be explained by the
different methods of measurement. While the mean propagation velocity
was calculated from the flame positions which were determined by the
contents of ashes in the grid cells, the flame surface area was
obtained by linear interpolation of the zero level-set of the
$G$-function in the cells. Additionally one has to keep in mind that 
(\ref{area-vel}) holds strictly only for a burning velocity that is
constant over the flame front. This is probably not fulfilled in our
simulation since we would expect an intrinsic Markstein-like behavior
as discussed above.

\subsection{Robustness of the numerical scheme}
\citet{reinecke1999a} reported difficulties with the implementation
of the complete version of the level set method. We examined
our implementation in order to evaluate the improvements that could be
achieved to
stabilize the algorithm. Robustness of the reconstruction
scheme is a prerequisite for its application to more complex
simulations such as flame evolution in presence of imprinted
turbulence or type Ia supernova explosions.

From given pre- and
post-front states which exactly
fulfill the Rankine-Hugoniot jump conditions \citet{reinecke1999a}
synthesize a mixed cell according to
(\ref{averagemass_eq})--(\ref{averagetotenergy_eq}) for a given
volume fraction of unburnt material $\alpha_0$. 
While trying to reconstruct for the burnt and unburnt states imposing
an unburnt volume fraction $\alpha$ that deviates slightly from the 
exact value $\alpha_0$, they find that the solver for the
nonlinear reconstruction equation system can enter values for which
the equation of state is undefined. It reaches values for the
internal energy of the unburnt material which are lower than the
internal energy belonging to zero temperature. This, of course means
that the reconstruction fails. The problem is caused by the
degenerate equation of state and is not observed in simulations of
chemical deflagration fronts. 

In simulations of thermonuclear flames this is a serious obstacle to
build a stable implementation, because the flame front is
linearly interpolated in mixed cells and therefore $\alpha$ deviates
from the exact value. This deviation becomes most pronounced in highly
curved front geometries. One would expect higher orders of
interpolation to cure this problem. This, however, would strongly
increase the number of topological uncertainties of the kind depicted
in the rightmost sketch of Figure \ref{alpha_fig}.

\begin{figure}[t!]
  \begin{center}
  \includegraphics[width=3.5cm]{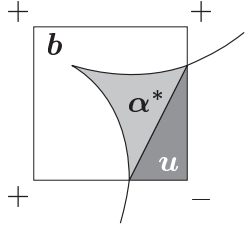}
  \end{center}
  \caption{Critical situation: the region marked with $\alpha^*$ is
  the deviation between the actual $\alpha_0$ and $\alpha$ determined
  by linear interpolation
  \label{crit}}
\end{figure}

The described difficulty remains a critical issue in our simulations. It occurs
when the front is highly curved. This is especially the case for cusps of
cellular stabilized front geometries. Here situations are possible
which lead to a large deviation of the $\alpha$-value determined by
linear interpolation from the intersection points of cell borders with
zero level set of $G$ and the actual $\alpha_0$ given by the exact
shape of the zero level line of $G$, as illustrated in Figure \ref{crit}. 
In these cases energy is added to the cell to obtain a physical
unburnt state with positive temperature. This of course makes our
numerical scheme slightly non-conservative. Fortunately, the cases
where cusps are aligned in an unfavorable manner with respect to the
computational grid are rare. Of course, as stated by
\citet{reinecke1999a}, a straight forward method to extenuate the
problem is to limit the flame curvature by introducing a
curvature-dependent effective burning velocity according to (\ref{markstein}).
The curvature $\kappa$ of the flame front can be
easily determined from $G$. In the presented simulations we did
not explicitly prescribe a burning law according to
(\ref{markstein}). For reasons discussed in {\S} \ref{sec_num_ld} a
Markstein-like behavior may be present for numerical reasons, but a
larger Markstein length might contribute to the robustness of the
in-cell reconstruction scheme.

A further issue in the \citet{reinecke1999a} implementation is
that the implicit treatment
of the source terms in the reconstruction scheme (see
{\S} \ref{in-cell_rec_sec}) was not applied. This leads to an
additional deviation of energy and species distribution from the zero
level set of the $G$-function. Implementation of this method in our
code made the reconstruction scheme more reliable.

\section{Conclusion}

As stated in the introduction, a thorough understanding of the flame
propagation in SN Ia on intermediate scales may be crucial to explain
these astrophysical events. To summarize the contribution of the
presented work to that issue it seems appropriate to repeat briefly
the current status of research in the area of thermonuclear flame
propagation in SN Ia in connection to the LD instability.

After the pioneering work of \citet{timmes1992a} which in detail
described the one-dimensional structure of thermonuclear flames in
white dwarf matter there have been several attempts to investigate the
propagation of two-dimensional flames. \citet{niemeyer1995a} were the
first to confirm by means of a hydrodynamical simulation that
thermonuclear flames in white dwarf matter are indeed subject to the
LD instability. They were, however, not able to reach the nonlinear
regime of flame propagation, where the flame is expected to stabilize
in a cellular shape. This phenomenon was subject to other
investigations, e.g. \cite{bychkov1995a} and
\cite{blinnikov1996a}. These studies are, however, based on the
investigation of the Sivashinsky equation and do not solve the full
hydrodynamical problem. Thus, although the assumption that the flame
is nonlinearly stabilized is supported by these investigations, it
can not be considered as proven.

With the presented numerical model we were able to confirm the statement of
\citet{niemeyer1995a} that nuclear burning fronts in white dwarfs are
subject to the LD instability on scales larger than the flame
width. The growth rate of perturbation amplitude in the linear regime
of the flame propagation is in reasonable agreement with the
theoretical prediction for sufficient spatial resolution. Beyond the
results of \citet{niemeyer1995a} we observe the transition from
linear to nonlinear propagation regime. The formation of cusps
pointing towards the burnt material stabilizes the flame in a cellular
shape. This has been observed and theoretically interpreted for
terrestrial flames and is now demonstrated by means of a full
hydrodynamical simulation for astrophysical flames for the first
time. 
With the increase of the flame surface we note a
flame acceleration which saturates at about 1.3 times the laminar
flame velocity when the flame has settled in its steady cellular
structure.

All our simulations with sufficient resolution finally yield a flame
in the shape of a single domain-filling cell and with higher
resolution this basic shape gets superimposed with small cells. This
in in excellent agreement with results from numerical simulations of the
Sivashinsky-equation. \citet{gutman1990a} study the flame evolution
according to this equation numerically using a simulation set-up
similar to ours. While their method is a semianalytical approach they
find effects which are very similar to our results. For domains only
a few times larger than the critical wavelength corresponding to the
maximum amplification rate they find a time-independent cusp-like
structure emerging. For larger domains (some ten times the critical
wavelength) they observe a time-dependent pattern of small-scale cells
superimposed to the cusp. They argue that this is caused by an
increased sensitivity of large-scale configurations at wide intervals
to small perturbations. Therefore the observed small-scale cells are
attributed to weak numerical noise rather than being a self-sustaining
phenomenon. 
We believe that this is the case in our simulations,
too. 
The observation of a similar behavior in our model
rules out the possibility that the effect is a peculiarity of the
numerical solution of the Sivashinsky equation.
Further discussion of the phenomenon and attempt for analytical
treatment using the pole decomposition method for solving a modified
Sivashinsky-equation can be found in \citet{joulin1989a} and
\citet{joulin1994a}.

Whether other geometries (like the spherically
expanding flame investigated by \citet{blinnikov1996a}) would lead to
a repeated cell-splitting needs further investigation.

\citet{niemeyer1995a} report a break-up of the cellular stabilization
at a fuel density of $5 \times 10^7 \mathrm{g}\,\mathrm{cm}^{-3}$. Our
simulations were performed at the same density of the unburnt material
and do not show this feature. This is, however, not necessarily a
contradiction since it is possible that perturbations as a result of numerical
noise, which is different in the two implementations, could trigger
the break-up. This scenario is not unrealistic since one would
expect physical noise to be present in SN Ia explosions.
The possibility of a loss of nonlinear
stabilization and self-turbulization of the flame front leading to
active turbulent combustion remains an open question that is of great
importance for large scale supernova simulations.
This topic will be investigated in a subsequent study aiming at
the simulation of the interaction of the LD instability with imprinted
turbulence. 
Here the flame is exposed to a ``controlled noise'', which can be quantified.
From the results of numerical simulations it is evident
that our model of thermonuclear flames is an adequate tool for this task.

\acknowledgments

The authors would like to thank M.~Reinecke, H.~Schmidt and R.~Klein
for their help with the numerical implementation and S.~Blinnikov for
stimulating discussions on theoretical aspects.

\end{document}